\documentclass[namedreferences]{kluwer}

\usepackage{epsfig}

\begin{document}
\begin{article}
\begin{opening}

\title{The Fossil Starburst in M82\thanks{Based on observations with
the NASA/ESA {\sl Hubble Space Telescope}, obtained at the Space
Telescope Science Institute, which is operated by the Association of
Universities for Research in Astronomy (AURA), Inc., under NASA contract
NAS 5-26555.}}

\author{Richard \surname{de Grijs}\email{grijs@virginia.edu}}
\author{Robert W. \surname{O'Connell}\email{rwo@virginia.edu}}
\institute{Astronomy Department, University of Virginia, P.O. Box 3818,
Charlottesville, VA 22903, USA}
\author{John S. \surname{Gallagher, III}\email{gallagher@astro.wisc.edu}}
\institute{Astronomy Department, University of Wisconsin, 475 North
Charter Street, Madison, WI 53706, USA}

\begin{abstract}
We present high-resolution optical and near-infrared {\sl HST}
observations of two adjacent regions in the fossil starburst region in
M82, M82 B1 and B2.  The presence of both the active and the fossil
starburst in M82 provides a unique physical environment to study the
stellar and dynamical evolution of star cluster systems.  \\
The cluster population in B2 is more heavily affected by internal
extinction than that in B1, amounting to an excess extinction in B2 of
$A_{V,{\rm excess}} \simeq 1.1\pm0.3$ mag.  Preliminary age estimates
date the cluster population in the fossil starburst between $\sim 2
\times 10^8$ and $\sim 10^9$ years.  The radial luminosity profiles of
the brightest clusters are more closely approximated by power laws than
by a Gaussian model, in particular in their wings, which favors a slow
star formation scenario. 
\end{abstract}


\end{opening}

\section{M82, the prototypical starburst galaxy}

Observations at all wavelengths from radio to X-rays are consistent with
a scenario that tidal interactions between M82 and another member galaxy
of the M81 group has channeled large amounts of gas into the central
regions of M82 during the last several 100 Myr (e.g., Telesco 1988,
Rieke et al.  1993).  This has induced a starburst which has continued
for up to about 50 Myr.  All of the bright radio and infrared sources
associated with the active starburst are confined to the galaxy's
center, lying within a radius of $\sim 500$ pc, and corresponding
spatially with bright optical structures, labeled M82 A, C, and E in
O'Connell \& Mangano (1978, OM78). 

However, evidence exists that this is not the only major starburst
episode to have occurred in M82.  A region at about 1 kpc NE from the
galactic center, M82 B (cf.  OM78), has exactly the properties one would
predict for a fossil starburst with a similar amplitude to the active
burst.  Marcum \& O'Connell (1996) find a sharp main-sequence cut-off of
the composite stellar system in M82 B, corresponding to an age of $\sim$
100--200 Myr and an average extinction of $A_V \sim 0.6$ mag.  Region A,
on the other hand, is only consistent with a very young population
($\sim 5$ Myr) and is more heavily affected by extinction ($A_V \sim
2.2$ mag).  By extrapolating region B's surface brightness ($\mu_V \sim
16.5$ mag arcsec$^{-2}$ [OM78], after correction for foreground
extinction) back to an age of 10 Myr we estimate that its surface
brightness was $\sim 2$ magnitudes brighter (cf.  the Bruzual \& Charlot
[1996, BC96] stellar population models), similar to that presently
observed in the active starburst.

{\sl HST} imaging of the bright central regions of M82 resolved these
into a swarm of young star cluster candidates, with a FWHM of $\sim 3.5$
pc ($0.''2$) and mean $L_V \sim 4 \times 10^6 L_\odot$ (O'Connell et al. 
1994, 1995), brighter than any globular cluster in the Local Group. 

This is the nearest rich system of such objects; such ``super star
clusters'' have also been discovered with {\sl HST} in other interacting
and amorphous systems, and in dwarf and starburst galaxies (e.g.,
Holtzman et al.  1992, Whitmore et al.  1993, O'Connell et al.  1994,
Conti et al.  1996, Ho 1997, Carlson et al.  1998, Watson et al.  1998,
among others).  Their diameters, luminosities, and -- in several cases
-- masses are consistent with these being young {\it globular} clusters
formed as a result of recent gas flows (e.g., van den Bergh 1995, Meurer
1995, Ho \& Filippenko 1996).  It is possible that most of the star
formation in starbursts takes place in the form of such concentrated
clusters. Our observations of M82 do not reveal similar cluster
formation outside the active and the fossil starburst regions.

Under the assumption that region B is indeed a fossil starburst site, it
is expected that it originally contained a complement of luminous
clusters similar to that now observed in region A.  The combination of
observations of both the active and the fossil starburst sites in M82
therefore provides a unique physical environment for the study of the
stellar and dynamical evolution of these star cluster systems. 

\section{{\sl HST} observations of the M82 central region}

The fossil starburst region, M82 B, was observed on September 15, 1997,
with both {\sl WFPC2} and {\sl NICMOS} on board the {\sl HST}.  We
imaged two adjacent $\sim35''$ square fields (Planetary Camera [PC]
field of view, $0.''0455$ pix$^{-1}$) in the M82 B region in the F439W,
F555W and F814W passbands, with total integration times of 4400s, 2500s
and 2200s, respectively, for region ``B1'', and 4100s, 3100s and 2200s,
respectively, for region ``B2''.  These observations were obtained using
four exposures per filter, covering a large range in integration times
to facilitate the removal of cosmic ray events.  The F439W, F555W and
F814W filters have roughly similar characteristics to the
Johnson-Cousins broad-band {\it B, V} and {\it I} filters, respectively. 

In the near-infrared (NIR) we chose to use {\sl NICMOS} Camera-2
($0.''075$ pix$^{-1}$), which provided the best compromise of resolution
and field of view.  We acquired 4 partially spatially overlapping
exposures in both the F110W and F160W filters (approximately similar to
the Bessell {\it J} and {\it H} filters, respectively) in a tiled
pattern; the integrations, of 512s each, were taken in MULTIACCUM mode
to preserve dynamic range and to correct for cosmic rays. 

\section{Selection procedure}

\subsection{Selection of real sources in a highly disturbed area}

Unfortunately, the separation of real sources from artefacts in M82 B is
problematic, due to significant small-scale variations in the amplitude
of the background emission, which are largely caused by the highly
variable extinction.  For this reason, we cannot use standard unsharp
masking techniques to remove this background, since this produces
significant residual emission along the dust features. 

An initial visual examination of the multi-passband observations
revealed a multitude of faint point sources that become increasingly
obvious with increasing wavelength.  To include in our source selection
a maximum number of real and a minimum amount of spurious sources (due
to, e.g., dust features, weak cosmic rays, or poisson noise in regions
of high surface brightness or at the CCD edges), we decided to cross
correlate source lists obtained in individual passbands.  We performed
extensive checks to find the best selection criterions and thus to
minimize the effects introduced by artefacts on the one hand and the
exclusion of either very red or very blue sources on the other. 

We chose our detection thresholds such that the number of candidate
sources selected from the images in all passbands were comparable, of
order 4000.  Then, we cross correlated the source lists obtained in the
individual passbands.  Finally, we determined which combination of
passbands resulted in the optimal matching of sources detected in both
the blue and the NIR passbands.  This source selection procedure lead us
to conclude that a final source list obtained from the cross correlation
of the candidate sources detected in the F555W and F814W filters would
contain the most representative fraction of the M82 B-region source
population. 

However, an initial visual examination of the cross-correlated {\it V}
and {\it I}-band sources showed that the automated detection routine had
returned non-detections as well as artefacts that were clearly
associated with dust features.  It also showed that the quality of the
resulting sources for photometric follow-up was highly variable. 
Therefore, we decided to check the reality of all cross-correlated
sources by examining them visually, in both passbands, at the same time
classifying them in terms of contrast, sharpness, and the presence of
nearby neighbors.  Moreover, we added (and verified) the $\sim$ 30\% of
the sources in each field that were missed by the automated detection
routine, but were clearly real sources, and were recognized as such by
the visual examination of both the {\it V} and the {\it I}-band images. 

The final source lists thus obtained contain 737 and 642 verified
sources in M82 B1 and B2, respectively. 

\subsection{Synthetic and observed star fields as control fields}

We estimated the completeness of our object list by randomly and
uniformly adding 500 synthetic point sources of input magnitudes between
20.0 and 25.0 mag to the observed images in both the F555W and the F814W
filters.  The PSFs of the synthetic point sources were obtained from
observational PSFs, and scaled to the desired magnitudes.  The effects
of crowding in our simulated star fields are small: only $\sim$ 1--2\%
of the simulated objects were not retrieved due to crowding or overlap
of adjacent sources. 

For the uniformly distributed simulated sources, we established the 50\%
completeness limits at F555W $> 23.1$ and 23.3 mag (for B1 and B2,
respectively) and at F814W $> 23.0$ mag. 

In addition to these synthetic star fields, we used our $\omega$Cen {\sl
HST} observations in the F555W passband, obtained as part of program
6053 (PI O'Connell, Cycle 5), to verify the reduction procedures and the
accuracy of our photometry. 

\subsection{Separating star clusters from stars}

We based the distinction between stars and more extended sources in the
M82 B fields on the statistical differences between the size
characteristics of the populations of real sources in these fields and
those of the stars in the $\omega$Cen control field.  To do so, we added
a scaled version of the $\omega$Cen field to the M82 B fields, such that
the output magnitudes of the majority of the $\omega$Cen stars were in
the same magnitude range as those of the M82 B sources, in the latter
reference frame. 

We determined characteristic sizes of both our M82 B verified sources
and the $\omega$Cen stars, using a Gaussian fitting routine.  Although
the true luminosity profiles of the star clusters in M82 B may differ
from Gaussians (Sect.  \ref{profiles.sect}), this method allows us to
distinguish between compact and extended sources.  A comparison between
the distribution of characteristic stellar sizes from the globular
cluster field with those of the verified sources in M82 B revealed that
the population of extended sources in M82 B is well-represented by
sources with $\sigma_{\rm Gaussian} \ge 1.25$ pixels.  Therefore, in the
following we will consider those sources with $\sigma_{\rm G} \ge 1.25$
and {\it V}-band magnitude brighter than the 50\% completeness limit to
be part of our verified cluster samples.  They contain 128 and 218
cluster candidates in B1 and B2, respectively. 

\section{Stellar population synthesis: tracing the evolution}

A close examination of the color distributions in both fields reveals
that the cluster population in B2 displays a peak at redder colors,
whereas the dispersion in colors is larger than in the B1 region, in
particular on the redward side.  Both these observations hint at a more
heavily obscured globular cluster population in B2.  Under the
assumption that the candidate cluster populations in B1 and B2 are
coeval, which is likely if they originated in the same starburst, the
differences between their color distributions can entirely be attributed
to extinction effects, amounting to an excess extinction in B2 of
$A_{V,\rm excess} \sim 1.1 \pm 0.3$ mag. 

Their color-magnitude diagrams (CMDs) reveal that the brightest young
cluster candidates have approximately identical colors and magnitudes in
either region: $(B-V)_0 \approx 0.6 \pm 0.2, (V-I)_0 \approx 1.0 \pm
0.2$ mag, and $17.5 \lesssim V \lesssim 19.5$; this indicates that these
are mainly unaffected by extinction. 

We take as tentative color limits for the determination of the intrinsic
age spread in the M82 B cluster population $0.2 \pm 0.1 \lesssim
(B-V)_{0,c} \lesssim 0.6 \pm 0.2$ and $0.4 \pm 0.2 \lesssim (V-I)_{0,c}
\lesssim 1.0 \pm 0.2$.  The limits on the blue side are set by the
colors of the bluest cluster candidates; the red limits are the colors
of the brightest (supposedly unobscured) clusters in either field, since
from this color range, most cluster colors can be achieved with up to
2(--3) magnitudes of visual extinction, which is not an unreasonably
large amount, considering the highly variable extinction features seen
in the images.  Employing the BC96 initial burst models yields ages
between $\sim 2 \times 10^8$ and $\sim 1 \times 10^9$ yr, assuming solar
metallicity.  From the optical--near-infrared color-color diagrams shown
in Fig.  1, we derive a lower limit to the allowed age range for the
bulk of the cluster population of 200--300 Myr, entirely consistent with
the age estimates obtained from the optical CMDs. 

\begin{figure}
\centerline{\epsfig{file=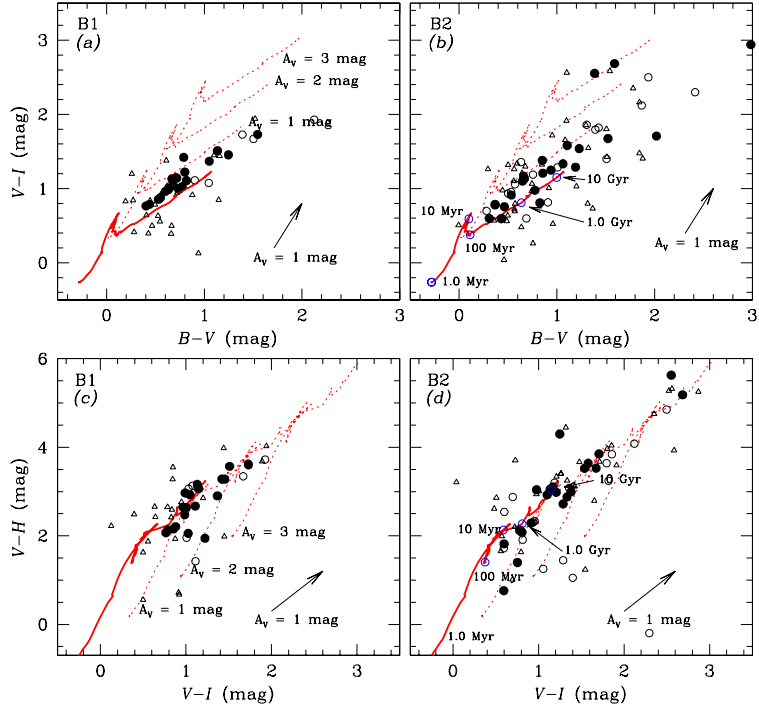,width=25pc}}
\caption{Optical and NIR color-color diagrams for the cluster candidates
in M82 B1 and B2, divided in magnitude bins (roughly corresponding to
quality bins); solid dots: $V \le 21.0$; open circles: $21.0 < V \le
22.0$; open triangles: $22.0 < V \le$ 50\% completeness limit. The
reddening vectors are shown as arrows; the effects of reddening on
unreddened evolutionary tracks of an instantaneous burst stellar
population (BC96, full drawn lines) are indicated by the dotted tracks.
We have only included those sources that were not significantly affected
by neirby neighbors or highly variable backgrounds.}
\end{figure}

\section{Cluster luminosity profiles in close-up}
\label{profiles.sect}

\begin{figure}
\centerline{\epsfig{file=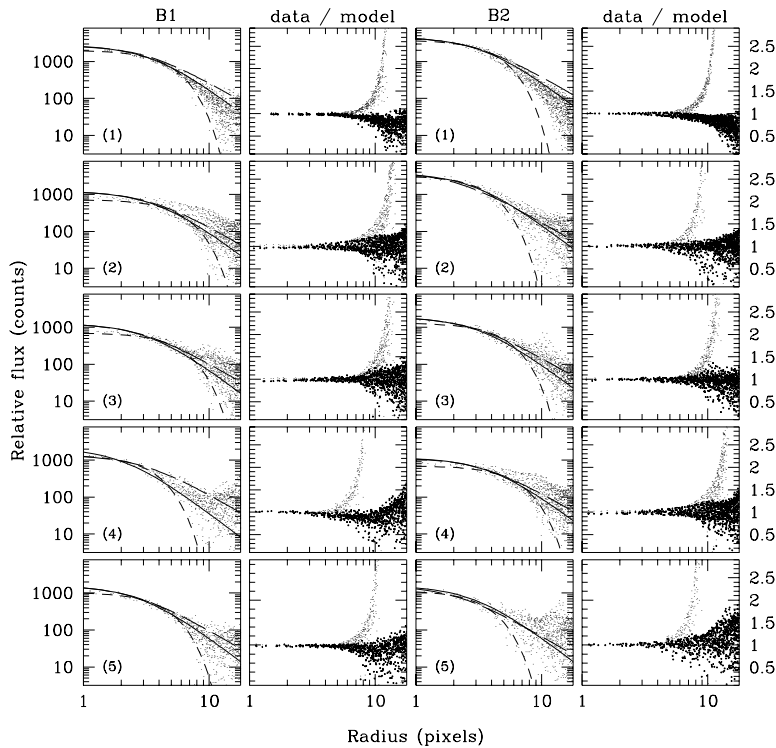,width=25pc}}
\caption{Luminosity profiles for the 5 brightest star clusters in both
B1 and B2.  Columns 1 and 3: data points, superposed with the
best-fitting Gaussian (short-dashed lines), modified Hubble ($\gamma =
2.0$, long-dashed lines), and $\gamma = 2.6$ profiles (Elson et al. 
1987).  Columns 2 and 4: ratios of data points to model fit -- dots:
Gaussian fits; crosses: modified Hubble profiles.  The ratio for the
$\gamma = 2.6$ (``Elson'') profiles is marginally to significantly
reduced compared to the modified Hubble profile ($\gamma = 2.0$) fits.}
\end{figure}

In Fig.  2 we present the light profiles of the five brightest star
cluster candidates in each field.  These objects have sufficient S/N
ratios and are relatively isolated, thus enabling us to follow the light
profiles out to reasonably large radii. 

Several cluster radial luminosity functions have recently been used in
the literature, the simplest of them being a two-dimensional Gaussian. 
Alternatively, more complex models have been used, that are better
representations of the true stellar light distribution in {\it globular}
clusters.  The most general of these is the mathematically convenient
function of surface brightness $\mu(r)$ as a function of radius {\it r},
proposed by Elson et al.  (1987):

\begin{equation}
\label{elson.eq}
\mu(r) = \mu_0 \Bigl( 1 + ( r / R_{\rm core} )^2 \Bigr)^{-\gamma/2} ,
\end{equation}
which reduces to a modified Hubble law for $\gamma = 2$.  In fact, Elson
et al.  (1987) found, for 10 rich star clusters in the LMC, that $2.2
\lesssim \gamma \lesssim 3.2$, with a median $\gamma = 2.6$. 

From these figures, it is obvious that even though all three fitting
functions may represent the inner profiles relatively well, the
discrepancy between the results in the outer parts is a strong argument
in favor of the more extended modified Hubble or $\gamma = 2.6$
(``Elson'') profiles. 

Elson et al.  (1987) try to explain the approximate power-law behavior
of the LMC cluster profiles at large radii in terms of different star
formation time scales.  They argue that the observations probably favor
slow star formation.  Under these circumstances {\em and} by assuming a
high star formation efficiency, relaxation of a sufficiently clumpy
gaseous protocluster through clump-clump or clump-star two-body
encounters may produce the observed power-law luminosity profiles. 

Some of the star clusters show evidence for asymmetrical structure,
which may mean that they contain subclumps, or have not yet reached
dynamical equilibrium.  A visual inspection of these clusters shows that
they are noticeably elongated.  In the case of these clusters in
particular, we may be witnessing ongoing merging processes.  However, to
strengthen this suggestion, we will need spectroscopic follow-up
observations.  Similar subclustering has been observed in the super star
clusters in NGC 1569 (O'Connell et al.  1994) and some of the rich LMC
star clusters (e.g., Fischer et al.  1993).

\end{article} 
\end{document}